\documentclass[aps, prl, reprint, superscriptaddress]{revtex4-1}
\usepackage[english]{babel}
\usepackage{amsmath,amsthm}
\usepackage{amsfonts}
\usepackage{xspace}
\usepackage{bm}
\usepackage{booktabs}
\usepackage{graphicx}
\usepackage{dcolumn}
\usepackage[mathlines]{lineno}
\usepackage[colorlinks]{hyperref}  
\usepackage{gensymb}
\usepackage{epstopdf}
\usepackage{newfloat}
\usepackage{color,soul}
\DeclareFloatingEnvironment[name={Figure S}]{suppfigure}
\DeclareFloatingEnvironment[name={S}]{suppEquation}

\newcommand*{\citen}[1]{%
  \begingroup
    \romannumeral-`\x 
    \setcitestyle{numbers}%
    \cite{#1}%
  \endgroup   
}

\begin{document}
\newcommand{\HOe}{\ensuremath{H_\mathrm{Oe}}\xspace}
\newcommand{\HOeTi}{\ensuremath{H_\mathrm{Oe,Ti}}\xspace}
\newcommand{\HOemax}{\ensuremath{H_\mathrm{Oe,max}}\xspace}
\newcommand{\Idc}{\ensuremath{I_\mathrm{dc}}\xspace}
\newcommand{\AlOx}{Al$_2$O$_3$}
\newcommand{\VPH}{\ensuremath{V_\mathrm{PH}}\xspace}
\newcommand{\DVPH}{\ensuremath{\Delta V_\mathrm{PH}}\xspace}
\newcommand{\Hx}{\ensuremath{H_\mathrm{x}}\xspace}
\newcommand{\Hy}{\ensuremath{H_\mathrm{y}}\xspace}
\newcommand{\HI}{\ensuremath{H_\mathrm{I}}\xspace}
\newcommand{\fTi}{\ensuremath{f_\mathrm{Ti}}\xspace}
\newcommand{\HFMR}{\ensuremath{H_\mathrm{FMR}}\xspace}
\newcommand{\HSO}{\ensuremath{H_\mathrm{SO}}\xspace}
\newcommand{\Vs}{\ensuremath{V_\mathrm{s}}\xspace}
\newcommand{\Va}{\ensuremath{V_\mathrm{a}}\xspace}
\newcommand{\Vmix}{\ensuremath{V_\mathrm{mix}}\xspace}
\newcommand{\tNiFe}{\ensuremath{t_\mathrm{NiFe}}\xspace}
\newcommand{\tTi}{\ensuremath{t_\mathrm{Ti}}\xspace}
\newcommand{\tCu}{\ensuremath{t_\mathrm{Cu}}\xspace}
\newcommand{\JNiFe}{\ensuremath{J_\mathrm{NiFe}}\xspace}
\newcommand{\JCu}{\ensuremath{J_\mathrm{Cu}}\xspace}
\newcommand{\HOeNM}{\ensuremath{H_\mathrm{Oe,NM}}\xspace}
\newcommand{\SiO}{SiO$_2$\xspace}
\newcommand{\aR}{\ensuremath{\alpha_\mathrm{R}}\xspace}
\newcommand{\muB}{\ensuremath{\mu_\mathrm{B}}\xspace}
\newcommand{\Ms}{\ensuremath{M_\mathrm{s}}\xspace}
\newcommand{\parallelsum}{\mathbin{\!/\mkern-5mu/\!}}


\title{Interfacial Spin-Orbit Torque without Bulk Spin-Orbit Coupling}%

\affiliation{ 
Department of Electrical and Computer Engineering, Northeastern University, Boston, MA 02115
}%

\author{Satoru Emori}%
\altaffiliation{ 
Current Address: Geballe Laboratory for Advanced Materials and Department of Applied Physics, Stanford University, Stanford, CA 94305 USA
}%
\email{
satorue@stanford.edu
}

\author{Tianxiang Nan}%
\altaffiliation{
Current Address: Department of Materials Science and Engineering, University of Wisconsin Madison, Madison, WI 53706 USA
}%

\author{Amine M. Belkessam}%

\author{Xinjun Wang}%

\author{Alexei D. Matyushov}%

\author{Christopher J. Babroski}%

\author{Yuan Gao}%

\author{Hwaider Lin}%

\author{Nian X. Sun}%

\date{February 10, 2016}

\begin{abstract}
An electric current in the presence of spin-orbit coupling can generate a spin accumulation that exerts torques on a nearby magnetization. 
We demonstrate that, even in the absence of materials with strong bulk spin-orbit coupling, a torque can arise solely due to interfacial spin-orbit coupling, namely Rashba-Eldestein effects at metal/insulator interfaces. 
In magnetically soft NiFe sandwiched between a weak spin-orbit metal (Ti) and insulator (\AlOx), this  torque appears as an effective field, which is significantly larger than the Oersted field and sensitive to insertion of an additional layer between NiFe and \AlOx.
Our findings point to new routes for tuning spin-orbit torques by engineering interfacial electric dipoles.
\end{abstract}
\maketitle

An electric current in a thin film with spin-orbit coupling can produce a spin accumulation~\cite{Edelstein1990, Hoffmann2013, Sinova2015}, which can then exert sizable torques on magnetic moments~\cite{Manchon2008,Gambardella2011a,Haney2013a,Brataas2014}.
First demonstrated in a ferromagnetic semiconductor~\cite{Chernyshov2009}, ``spin-orbit torques'' are nowadays studied in room-temperature ferromagnetic metals (FMs) interfaced with heavy metals (HMs) with strong spin-orbit coupling, such as Pt, Ta, and W~\cite{Liu2012d, Fan2013, Fan2014, Pai2014, Miron2011, Skinner2014, Kawaguchi2015, Allen2015, Garello2013, Kim2013a, Avci2014, Liu2014b, Emori2014c, Qiu2015, Akyol2015, Sato2016}. 
These torques can arise from (1) spin-dependent scattering of conduction electrons in the bulk of the HM, i.e., the spin-Hall effect~\cite{Hoffmann2013, Sinova2015, Liu2012d, Fan2013, Fan2014, Pai2014}, and (2) momentum-dependent spin polarization at the HM/FM interface, i.e., the Rashba-Edelstein effect~\cite{Edelstein1990, Gambardella2011a, Miron2011, Skinner2014, Kawaguchi2015, Allen2015}. 
Since a HM/FM system can exhibit either or both of these spin-orbit effects, it can be a challenge to distinguish the spin-Hall and Rashba-Edelstein contributions~\cite{Garello2013, Kim2013a, Haney2013a,Brataas2014, Sinova2015}. 
Spin-orbit torques may be further influenced by spin scattering~\cite{Chen2015b} or proximity-induced magnetization~\cite{Zhang2015c} at the HM/FM interface. 
Moreover, in many cases~\cite{Miron2011, Liu2012d, Skinner2014, Kawaguchi2015, Allen2015, Fan2013, Fan2014, Pai2014, Avci2014, Garello2013, Kim2013a, Liu2014b, Emori2014c, Qiu2015, Akyol2015, Sato2016}, the FM interfaced on one side with a HM is interfaced on the other with an insulating material, and the electric dipole at the FM/insulator interface~\cite{Xu2012, Ibrahim2016} may also give rise to a Rashba-Edelstein effect. 
Recent studies~\cite{Liu2014b, Emori2014c, Qiu2015, Akyol2015, Sato2016} indeed suggest nontrivial influences from insulating-oxide capping layers in perpendicularly-magnetized HM/FM systems.
However, with the FM only $\lesssim$1 nm thick~\cite{Liu2014b, Emori2014c, Qiu2015, Akyol2015, Sato2016}, changing the composition of the capping layer may modify the ultrathin FM and hence the HM/FM interface.
The points above make it difficult to disentangle the contributions from the HM bulk, HM/FM interface, and FM/insulator interface, thereby posing a challenge for coherent engineering of spin-orbit torques.

In this Letter, we experimentally show a spin-orbit torque that emerges exclusively from metal/insulator interfaces in the absence of materials with strong bulk spin-orbit coupling. 
Our samples consist of magnetically soft Ni$_{80}$Fe$_{20}$ (NiFe) sandwiched between a weak spin-orbit light metal (Ti) and a weak spin-orbit insulator (\AlOx). 
We observe a ``field-like'' spin-orbit torque that appears as a current-induced effective field, which is significantly larger than the Oersted field.
This torque is conclusively attributed to the Rashba-Edelstein effect, i.e., spin accumulation at the NiFe/\AlOx\ interface exchange coupling to the magnetization in NiFe~\cite{Manchon2008, Gambardella2011a}. 
We also observe a ``nonlocal'' torque with Cu inserted between NiFe and \AlOx\ due to spin accumulation at the Cu/\AlOx\ interface. 
Our findings demonstrate simple systems exhibiting purely interfacial spin-orbit coupling, which are free from complications caused by strong spin-orbit HMs, and open possibilities for spin-orbit torques enabled by engineered electric dipoles at interfaces. 

Thin-film heterostructures are sputter-deposited on Si substrates with a 50-nm thick SiO$_2$ overlayer.  
All layers are deposited at an Ar pressure of $3\times10^{-3}$ Torr with a background pressure of $\lesssim$2$\times10^{-7}$ Torr. 
Metallic layers are deposited by dc magnetron sputtering, whereas \AlOx\ is deposited by rf magnetron sputtering from a compositional target. 
The deposition rates are calibrated by X-ray reflectivity. 
For each structure, unless otherwise noted, a 1.2-nm thick Ti seed layer is used to promote the growth of NiFe with narrower resonance linewidth and near-bulk saturation magnetization. 
Devices are patterned and contacted by Cr(3 nm)/Au(100 nm) electrodes by photolithography and liftoff. 

\begin{figure}[tb]
\includegraphics[width=1.0\columnwidth]{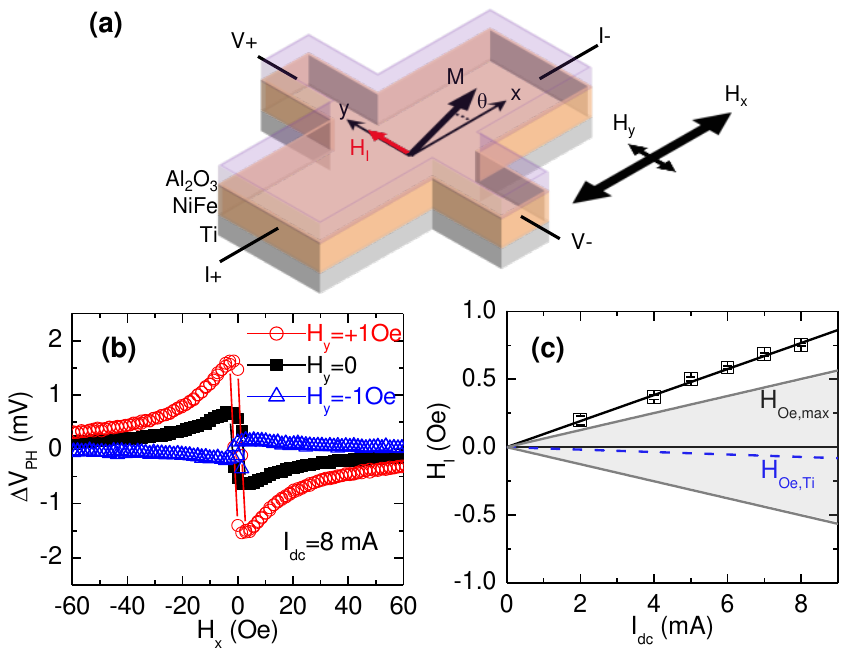}
\caption{\label{fig:PHE} (a) Schematic of the second-order PHE measurement. 
(b) Second-order planar Hall voltage \DVPH curves at different transverse bias fields \Hy. 
(c) Current-induced field \HI versus \Idc. The dotted line shows \HOeTi based on the estimated fraction of \Idc in Ti. The shaded area is bounded by the maximum possible Oersted field \HOemax. 
} 
\end{figure}

We first examine the current-induced field in a trilayer of Ti(1.2 nm)/NiFe(2.5 nm)/Al$_2$O$_3$(1.5 nm) by using the second-order planar Hall effect (PHE) voltage technique devised by Fan \textit{et al.}~\cite{Fan2013, Fan2014}. 
As illustrated in Fig.~\ref{fig:PHE}(a), a dc current \Idc along the x-axis generates a planar Hall voltage \VPH along the y-axis in a 100-$\mu$m wide Hall bar, which is placed in the center of a two-axis Helmholtz coil.
The second-order planar Hall voltage $\DVPH =  \VPH(+\Idc)+\VPH(-\Idc)$ is measured while sweeping the external field \Hx (Fig.~\ref{fig:PHE}(b)).  
The total current-induced in-plane transverse field \HI (which includes the Oersted field) pulls the magnetization away from the x-axis at an angle $\theta$.   
When $|\Hx|$ is sufficiently large ($\gtrsim$10 Oe), $\theta$ is small and \DVPH is proportional to $\Idc^2\Hx^{-1} d\HI/d\Idc$~\cite{Fan2013}. 
Following the procedure in Ref.~\citen{Fan2014}, we apply a constant transverse bias field $|\Hy|=1$ Oe (Fig.~\ref{fig:PHE}(a),(b)) and extrapolate the critical \Hy required to cancel \HI, i.e., to null the \DVPH spectrum. 
For the data in Fig.~\ref{fig:PHE}(b), \Hy=-0.75 Oe would null \DVPH, so \HI = 0.75 Oe at \Idc = 8 mA.  

As shown in Fig.~\ref{fig:PHE}(c), \HI scales linearly with \Idc with slope d\HI/d\Idc = 0.095 Oe per mA.  
To estimate the Oersted field contribution to \HI, the current is assumed to be uniform within each conductive layer, such that the Oersted field comes only from the current in the Ti layer, $\HOeTi = \fTi\Idc/2w$, where \fTi is the fraction of \Idc in Ti and $w$ is the Hall bar width.  
The sheet resistances 2000 $\Omega$/sq for Ti(1.2 nm)  and 350 $\Omega$/sq for NiFe(2.5 nm), found from four-point resistance measurements, yield \fTi = 0.15 and $|\HOeTi| = 0.009$ Oe per mA.  
The net \HI is therefore an order of magnitude larger than \HOeTi, and moreover, the direction of \HI opposes \HOeTi.  

The actual Oersted field may deviate from \HOeTi because of nonuniform current distribution within each conductive layer and interfacial scattering, both of which are difficult to quantify.  
However, we can place the upper bound on the Oersted field, $|\HOemax| = |\Idc|/2w$, by assuming that the \emph{entire} \Idc flows above or below the magnetic layer.     
In Fig.~\ref{fig:PHE}(c), we shade the range bounded by $|\HOemax|$.  
The magnitude of \HI still exceeds \HOemax, confirming the presence of an additional current-induced field with a component collinear with the Oersted field.  

\begin{figure}[tb]
\includegraphics[width=1.0\columnwidth]{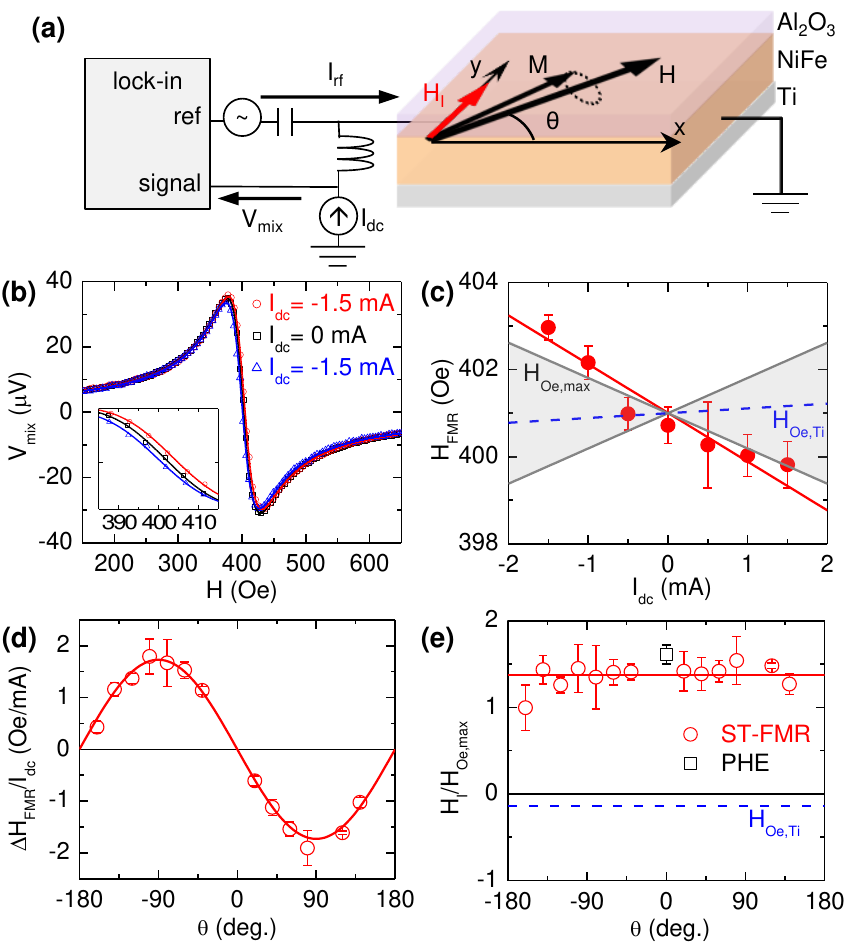}
\caption{\label{fig:STFMR} (a) Schematic of the ST-FMR setup. (b) ST-FMR spectra at different dc bias currents \Idc, 
 with rf current excitation at 5 GHz and +8 dBm and external field $H$ at $\theta = 40^{\circ}$. Inset: \Idc-induced shift of ST-FMR spectra. 
(c) Shift of resonance field \HFMR due to \Idc at $\theta = 40^{\circ}$. The error bar is the standard deviation of 5 measurements. The dotted line shows the estimated Oersted field from Ti, \HOeTi. The shaded area is bounded by the maximum possible Oersted field, \HOemax. 
(d)~Angular dependence of \Idc-induced \HFMR shift. The solid curve indicates the fit to $\sin\theta$. (e) Transverse current-induced field $\HI=-\Delta \HFMR/\sin\theta$ normalized by $\HOemax$ at various $\theta$.  The error bar is the error in linear fit of \HFMR versus \Idc. The solid line indicates the average of the ST-FMR data points.  The dotted line indicates estimated \HOeTi. The PHE data point at $\theta=0$ is the average of three devices.} 
\end{figure}

We also measure \HI with a technique based on spin-torque ferromagnetic resonance (ST-FMR)~\cite{Liu2011, Fang2011}.  
As illustrated in Fig.~\ref{fig:STFMR}(a), the rf excitation current is injected into a 5-$\mu$m wide, 25-$\mu$m long strip through a ground-signal-ground electrode.  
While the in-plane external field $H$ is swept at an in-plane angle $\theta$, the rectified mixing voltage \Vmix across the strip is acquired with a lock-in amplifier.
The resulting spectrum is well fit to a Lorentzian curve $\Vmix = \Vs F_\mathrm{s} + \Va F_\mathrm{a}$ consisting of the symmetric component $F_\mathrm{s} = W^2/((H-\HFMR)^2+W^2)$ and antisymmetric component $F_\mathrm{a} = W(H-\HFMR)/((H-\HFMR)^2+W^2)$, where $W$ is the resonance linewidth and \HFMR is the resonance field.  
We inject a small dc bias current $|\Idc|$$\leq$2 mA to measure the shift in \HFMR caused by the net \Idc-induced field \HI~\citen{Nan2015a}. 
Although the scatter in the ST-FMR data is greater than the PHE data (Fig.~\ref{fig:PHE}(c)), Fig.~\ref{fig:STFMR}(c) shows that the observed shift in \HFMR is significantly larger than (and opposes) the contribution from \HOeTi, and its magnitude exceeds the maximum possible shift from \HOemax.  

Fig.~\ref{fig:STFMR}(d) shows the \Idc-induced shift $\Delta \HFMR$ as a function of in-plane magnetization angle, equal to the applied field angle $\theta$ for the soft NiFe layer.  
This angular dependence is well described by a $\sin\theta$ relation, which implies that \HI is transverse to the current axis.
Fig.~\ref{fig:STFMR}(e) shows that the constant $\HI = -\Delta \HFMR/\sin\theta$ indeed agrees well with the PHE data measured at $\theta\approx0$. 
This finding confirms that \HI, including the non-Oersted contribution, is entirely transverse to the current and is independent of the magnetization orientation. 

\begin{figure}[tb]
\includegraphics[width=1.0\columnwidth]{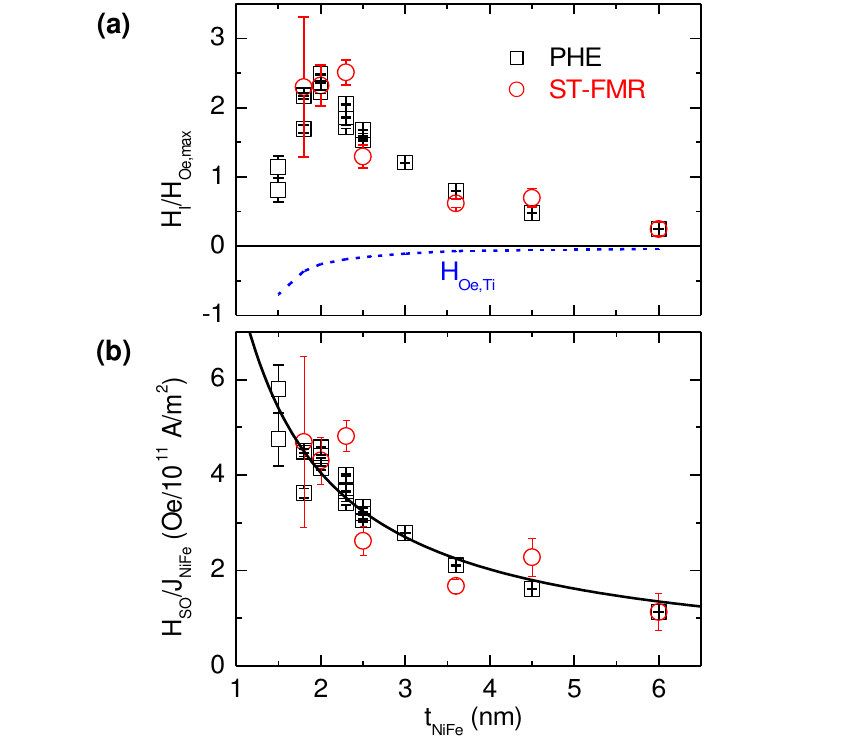}
\caption{\label{fig:NiFe} (a) NiFe-thickness \tNiFe dependence of \HI normalized by \HOemax. The dotted curve indicates the estimated Oersted field from Ti, \HOeTi. Each ST-FMR data point is the mean of results at several frequencies 4-7 GHz at $\theta = 45^{\circ}$ and $-135^{\circ}$.
$\HI/\HOemax > 0$ is defined as $\HI\parallelsum+y$ when $\Idc\parallelsum+x$ (illustrated in Figs.~\ref{fig:PHE}(a) and \ref{fig:STFMR}(a)).
(b) Estimated spin-orbit field \HSO per unit current density in NiFe, \JNiFe. The solid curve indicates the fit to ${\tNiFe}^{-1}$.} 
\end{figure}

For a wide range of NiFe thickness \tNiFe, as shown in Fig.~\ref{fig:NiFe}(a), we observe \HI that cannot be accounted for by the Oersted field alone. 
The observed \HI opposes \HOeTi in all samples, and \HI is more than a factor of 2 larger than \HOemax at $\tNiFe\approx 2$ nm. 
The drop in \HI for $\tNiFe \lesssim 2$ nm is caused by the increasing magnitude of \HOeTi, as NiFe becomes more resistive and a larger fraction of current flows through Ti with decreasing \tNiFe. 

The anomalous portion of \HI, which cannot be explained by the classical Oersted field, may be due to a spin-orbit torque that acts as a ``spin-orbit field'' \HSO. 
In Fig.~\ref{fig:NiFe}(b), we plot the estimated $\HSO = \HI - \HOeTi$ normalized by the current density in NiFe, \JNiFe.
This normalized \HSO scales inversely with \tNiFe, implying that the source of \HSO is outside or at a surface of the NiFe layer.
Therefore, \HSO does not arise from spin-orbit effects within the bulk of NiFe~\cite{Taniguchi2015}, i.e., the reciprocal of the recently reported inverse spin-Hall effect in FMs~\cite{Miao2013, Tsukahara2014, Azevedo2014, Wang2014e}. 
Moreover, any possible spin-orbit toques arising from the bulk of NiFe would depend on the magnetization orientation~\cite{Taniguchi2015} and are thus incompatible with the observed symmetry of \HSO (Fig.~\ref{fig:STFMR}(e)). 
It is unlikely that \HSO is generated by the spin-Hall effect in Ti, because its spin-Hall angle is small ($<$0.001)~\cite{Du2014b, Uchida2014a} and only a small fraction of \Idc is expected to be in the resistive ultrathin Ti layer.  
In Ti/NiFe/\AlOx, we also do not observe a damping-like torque that would be expected to arise from the spin-Hall effect~\cite{Haney2013a, Freimuth2014}; the linewidth $W$ is invariant with \Idc within our experimental resolution $<$0.2 Oe/mA~\cite{Nan2015a}.  

With spin-orbit effects in the bulk of NiFe and Ti ruled out as mechanisms behind \HSO, the only known mechanism that agrees with the observed \HSO is the Rashba-Edelstein effect~\cite{Edelstein1990, Manchon2008, Gambardella2011a}:
an interfacial spin accumulation (polarized transverse to the current) exchange couples to the magnetization in NiFe.
Indeed, tight-binding Rashba model calculations reveal a field-like torque, but no damping-like torque, in the first order of spin-orbit coupling due to transverse spin accumulation independent of the magnetization orientation~\cite{Kalitsov2016}.

\begin{figure}[tb]
\includegraphics[width=1.0\columnwidth]{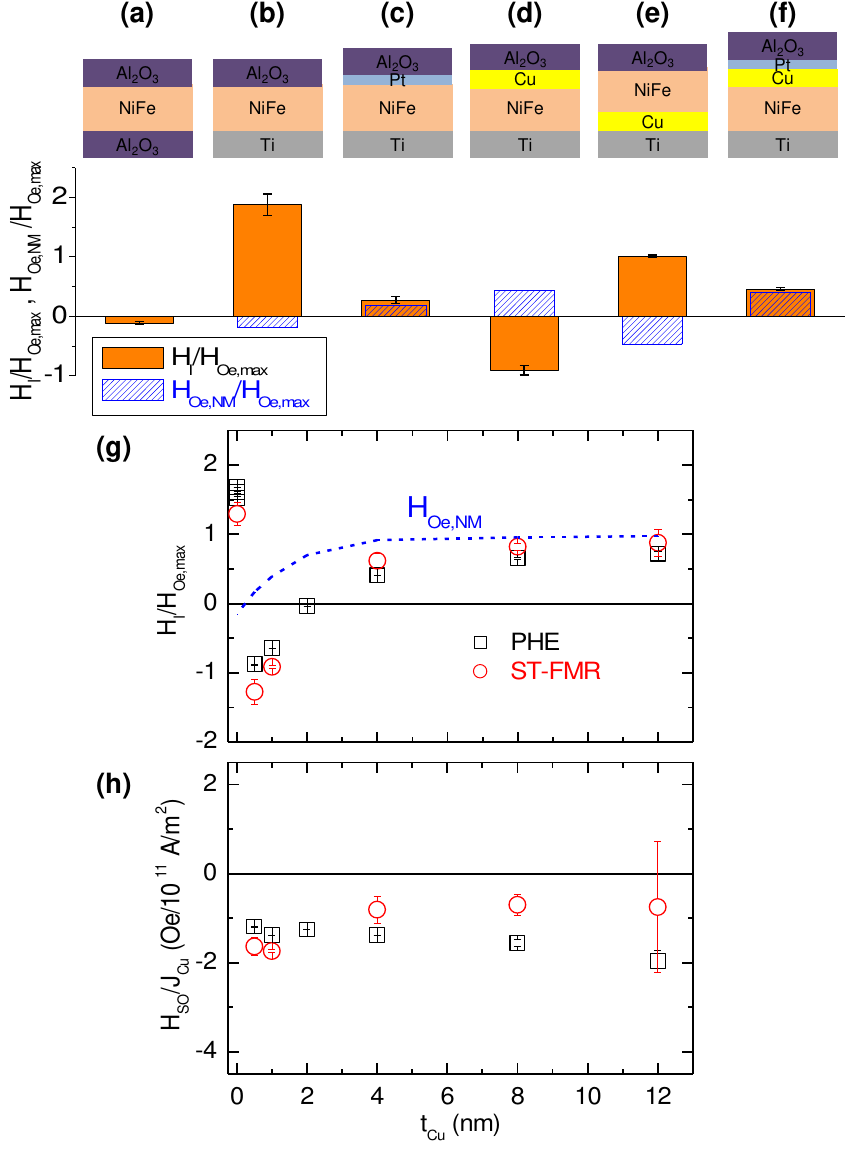}
\caption{\label{fig:layers} (a-f) Structural dependence of \HI (mean of measurements on three PHE devices) normalized by \HOemax. \HOeNM is the Oersted field from current in the nonmagnetic metal layers (Ti, Cu, Pt). The nominal layer thicknesses are NiFe: 2.3 nm, \AlOx: 1.5 nm, Ti: 1.2 nm, Cu: 1.0 nm, and Pt: 0.5 nm. 
(g) Cu-thickness \tCu dependence of \HI normalized by \HOemax at NiFe thickness 2.5 nm.  The blue dotted curve indicates \HOeNM. 
(h) Estimated spin-orbit field \HSO per unit current density in Cu, \JCu.} 
\end{figure}

We now gain further insight into the origin of \HSO by examining its dependence on the layer stack structure, as summarized in Fig.~\ref{fig:layers}(a-f).
In the symmetric \AlOx(1.5 nm)/NiFe(2.3 nm)/\AlOx(1.5 nm) trilayer (Fig.~\ref{fig:layers}(a)), \HI vanishes, which is as expected because the Oersted field should be nearly zero and the two nominally identical interfaces sandwiching NiFe produces no net spin accumulation.  
Breaking structural inversion symmetry with the Ti(1.2 nm) seed layer results in an uncompensated interfacial spin accumulation that generates a finite $\HSO=\HI-\HOeTi$ (Fig.~\ref{fig:layers}(b)). 

Inserting Pt(0.5 nm) between the NiFe and \AlOx\ layers suppresses \HSO, such that the estimated Oersted field \HOeNM from the nonmagnetic Ti and Pt layers entirely accounts for \HI (Fig.~\ref{fig:layers}(c)).
This may seem counterintuitive since Pt exhibits strong spin-orbit coupling and a large Rashba-Edelstein effect may be expected at the Pt surface~\cite{Zhang2014}.   
However, Pt is also a strong spin scatterer, as evidenced by an increase in the Gilbert damping parameter from $\approx$0.013 for Ti/NiFe/\AlOx\ to $\approx$0.03 for Ti/NiFe/Pt/\AlOx.
The accumulated spins may quickly become scattered by Pt, such that there is no net field-like torque mediated by exchange coupling~\cite{Manchon2008, Gambardella2011a} between these spins and the magnetization in NiFe.  
Based on the suppression of \HSO by Pt insertion, we infer that the Rashba-Edelstein effect at the NiFe/\AlOx\ interface is the source of \HSO. 

Inserting Cu(1 nm) at the NiFe/\AlOx\ interface \emph{reverses} the direction of $\HSO = \HI - \HOeNM$ (Fig.~\ref{fig:layers}(d)). 
We deduce a Rashba-Edelstein effect (opposite in sign to that of NiFe/\AlOx) at the Cu/\AlOx\ interface, rather than the NiFe/Cu interface, because
(1) if NiFe/Cu generates the reversed \HSO, we should see an enhanced \HSO for NiFe sandwiched between Cu (bottom) and \AlOx\ (top), but this is not the case (Fig.~\ref{fig:layers}(e));  
and (2) inserting a spin-scattering layer of Pt(0.5 nm) between Cu and \AlOx\ suppresses \HSO (Fig.~\ref{fig:layers}(f)). 

Fig.~\ref{fig:layers}(g) plots the dependence of \HI on Cu thickness \tCu. 
In the limit of large \tCu ($\approx$10 nm), \HI approaches \HOeNM that is predominantly due to the current in the highly conductive Cu layer. 
From the estimated current distribution, we obtain $\HSO = \HI-\HOeNM$ normalized by the current density in the Cu layer, \JCu. 
As shown in Fig.~\ref{fig:layers}(h), $\HSO/\JCu \approx$ 1-2 Oe/10$^{11}$ A/m$^2$ exhibits little dependence on \tCu.
This is consistent with the Rashba-Edelstein effect at the Cu/\AlOx\ interface that is present irrespective of \tCu.  
Persistence of \HSO even at large \tCu implies a nonlocal Rashba-Edelstein field: the spin accumulation at the Cu/\AlOx\ interface couples to the magnetization in NiFe across the Cu layer.
However, further studies are required to elucidate the mechanism involving Cu, since we do not observe any apparent oscillation in \HSO with \tCu that would be expected for exchange coupling across Cu~\cite{Parkin1991}. 

At $\tCu \approx 2$ nm, \HI vanishes because \HSO and \HOeNM compensate each other (Fig.~\ref{fig:layers}(g)). 
Fan \textit{et al}. also show near vanishing of \HI in NiFe(2 nm)/Cu(\tCu)/SOi$_2$(3.5 nm) at $\tCu \approx3$ nm~\cite{Fan2013}, and Avci \textit{et al}. report a current-induced field in Co(2.5 nm)/Cu(6 nm)/AlO$_\mathrm{x}$(1 nm) that is well below the estimated Oersted field~\cite{Avci2014}.
In each of these studies~\cite{Fan2013, Avci2014}, a spin-orbit field due to the Rashba-Edelstein effect at the Cu/oxide interface may have counteracted the Oersted field.

In summary, we have shown a current-induced spin-orbit torque due to Rashba-Edelstein effects at NiFe/\AlOx\ and Cu/\AlOx\ interfaces. 
This torque is distinct from previously reported spin-orbit torques because it arises even without spin-orbit coupling in the bulk of the constituent materials.
The origin of this torque is purely interfacial spin-orbit coupling, which likely emerges from the electric dipoles that develop at the metal/insulator interfaces~\cite{Xu2012, Ibrahim2016}. 
This mechanism is supported by recent theoretical predictions of current-induced spin polarization at metal/insulator interfaces in the absence of bulk spin-orbit coupling~\cite{Wang2013c, Borge2014}. 
Rashba-Edelstein effects at metal/insulator interfaces may be universal and should motivate the use of various previously-neglected materials as model systems for interfacial spin-dependent physics and as components for enhancing spin-orbit torques, perhaps combined with gate-voltage tuning~\cite{Liu2014b, Emori2014c, Bauer2015}.  
One possibility is to apply interfacial band alignment techniques, similar to those for semiconductor heterostructures~\cite{Kroemer2001}, to engineer giant dipole-induced Rashba-Edelstein effects. 
\\
\\
This work was supported by the AFRL through contract FA8650-14-C-5706, the W.M. Keck Foundation, and the NSF TANMS ERC Award 1160504.
X-ray reflectivity was performed in CMSE at MIT, and lithography was performed in the George J. Kostas Nanoscale Technology and Manufacturing Research Center.
We thank Geoffrey Beach, Carl Boone, Xin Fan, Adrian Feiguin, Chi-Feng Pai, and Kohei Ueda for helpful discussions.   
We give special thanks to Mairbek Chshiev, Sergey Nikolaev, and Noriyuki Sato for their comments and sharing of unpublished results.

\end{document}